\documentclass[pra,aps,preprint,showpacs,tightenlines,unsortedaddress,superscriptaddress]{revtex4}
\usepackage{graphics,bm}
\begin{document}

\title{Noisy Dynamics of a Vortex in a Partially Bose-Einstein Condensed Gas}

\author{R.A. Duine}
\email{duine@phys.uu.nl} \homepage{http://www.phys.uu.nl/~duine}
\affiliation{Institute for Theoretical Physics,
         University of Utrecht, Leuvenlaan 4,
         3584 CE Utrecht, The Netherlands}
\author{B.W.A. Leurs}
\email{leurs@lorentz.leidenuniv.nl}
\affiliation{Instituut-Lorentz for Theoretical Physics,
University of Leiden, Niels Bohrweg 2, 2333 CA Leiden, The
Netherlands}
\author{H.T.C. Stoof}
\email{stoof@phys.uu.nl} \homepage{http://www.phys.uu.nl/~stoof}
\affiliation{Institute for Theoretical Physics,
         University of Utrecht, Leuvenlaan 4,
         3584 CE Utrecht, The Netherlands}
\date{\today}

\begin{abstract}
We study the dynamics of a straight vortex line in a partially
Bose-Einstein condensed atomic gas. Using a variational approach
to the stochastic field equation that describes the dynamics of
the condensate at nonzero temperature, we derive the stochastic
equations of motion for the position of the vortex core. Using
these results, we calculate the time it takes the vortex to spiral
out of the condensate. Due to the fact that we include thermal
fluctuations in our description, this lifetime of the vortex is
finite even if its initial position is in the center of the
condensate.
\end{abstract}


\maketitle

% definitions
\def\bx{{\bf x}}
\def\bk{{\bf k}}
\def\half{\frac{1}{2}}
\def\args{(\bx,t)}

\section{Introduction}
\label{sec:intro} Contrary to classical fluids, superfluids
support rotation only through quantized vortices. Since quantized
vortices are therefore one of the hallmarks of superfluidity, the
experimental and theoretical study of these topological
excitations has attracted a great deal of attention in the field
of Bose-Einstein condensed gases. Following their first
experimental observation \cite{matthews1999}, they have now been
observed and experimentally studied by various groups
\cite{madison2000,raman2001,hodby2002}.

Theoretically, the dynamics of a single vortex line in a
Bose-Einstein condensate has been studied extensively in the
zero-temperature limit \cite{fetter2001}. In the absence of
external rotation, the vortex is predicted to precess around the
center of the condensate, which has indeed been observed
experimentally by Anderson {\it et al.} \cite{anderson2000}.
However, in this experiment it is also observed that the distance
of the vortex core to the center of the condensate increases with
time, i.e., the vortex spirals out of the condensate. This
observation is a sign of the presence of dissipation and thus
cannot be explained on the basis of a zero-temperature approach.
To understand it, we have to include the effects of the
noncondensed component of the gas. Although the effects of the
noncondensed thermal cloud on the equilibrium properties of
rotating Bose gases have been investigated
\cite{stringari1999,virtanen2001,mizushima2001,isoshima2003}, the
nonequilibrium dynamics of a single vortex at nonzero temperatures
has attracted relatively little attention.

It is the purpose of this paper to study the effects of the
thermal cloud on the motion of a single vortex in a Bose-Einstein
condensate. The starting point of our study is the stochastic
Gross-Pitaevskii equation derived by one of us
\cite{stoof1997,stoof1999}. This equation generalizes the usual
Gross-Pitaevskii equation, that provides an accurate description
of the dynamics of the condensate at zero temperature, to nonzero
temperatures by the inclusion of a dissipative term that describes
the growth or decay of the condensate. Moreover, thermal
fluctuations are included by an additive noise term that is
related to the growth or decay of the condensate by means of a
fluctuation-dissipation theorem. Although this equation can be
solved numerically \cite{stoof2001}, we intend to capture as much
of the physics as possible by using a variational approach to
stochastic field equations that was developed by two of us
\cite{duine2001}.

To make the paper more self-contained, we briefly discuss the
zero-temperature dynamics of a vortex in a Bose-Einstein
condensate in Sec.~\ref{sec:zerot}. In Sec.~\ref{sec:finitet} we
then derive the stochastic equations of motion that describe the
motion of the vortex core at a nonzero temperature. In
Sec.~\ref{sec:lifetime} we use these equations to derive an
equation for the average distance of the vortex to the center of
the condensate and use the latter result to calculate the lifetime
of the vortex. We compare our results with the available
theoretical results \cite{fedichev1999, schmidt2003}.
Unfortunately, there does not exist a detailed experimental study
of the lifetime of the vortex. We end in
Sec.~\ref{sec:conclusions} with our conclusions.

\section{Vortex dynamics at zero temperature} \label{sec:zerot}
The dynamics of a Bose-Einstein condensate is, at zero
temperature, well described by the Gross-Pitaevskii equation
\begin{equation}
\label{eq:gpeqn}
  i \hbar \frac{\partial \phi \args}{\partial t}
   = \left[
 -\frac{\hbar^2 {\bf \nabla}^2}{2m} + V^{\rm ext} (\bx) +
 T^{\rm 2B}|\phi \args|^2 \right] \phi \args~,
\end{equation}
for its macroscopic wavefunction $\phi \args$. Here, $m$ is the
atomic mass, and $T^{\rm 2B} = 4 \pi a \hbar^2/m$ is the two-body
T(ransition) matrix at zero energy, with $a>0$ the $s$-wave
scattering length of the atoms of interest. We take the external
trapping potential of the form
\begin{equation}
  V^{\rm ext} (\bx) = \frac{1}{2} m \left[ \omega^2 (x^2+y^2) +
  \omega_z^2 z^2 \right]~,
\end{equation}
with $\omega_z \gg \omega$, which implies that we have a
pancake-shaped condensate. The reason for choosing this geometry
is that it allows us to neglect the curvature of the vortex line,
provided that the vortex is close to the center of the condensate.

To study the dynamics of a vortex at zero temperature we use the
variational {\it ansatz}
\begin{eqnarray}
\label{eq:ansatzzerot}
  \phi \args &=& \frac{\sqrt{N_{\rm c}}}{\sqrt{\pi^{3/2} q^2 q_z \left[q^2+u_x^2 (t)+u^2_y(t)\right]}}
  \left\{\left[x-u_x(t)\right] +i \left[y-u_y(t)\right] \right\}
  \nonumber \\ && \times \exp \left\{ -\frac{x^2+y^2}{2 q^2} - \frac{z^2}{2
  q_z^2}\right\}~,
\end{eqnarray}
where $N_{\rm c}$ is the number of condensate atoms. This {\it
ansatz} has the same form as the exact wave function for a
noninteracting Bose-Einstein condensate with a vortex along the
$z$-axis. We treat the width of the condensate in the radial and
axial directions, denoted by $q$ and $q_z$, respectively, as
time-independent variational parameters. The pancake-shaped
geometry implies that $q_z \ll q$. The coordinates of the vortex
in the $x$-$y$~plane, denoted by $u_i (t)$, are taken to be
time-dependent variational parameters. For the case that the
vortex is close to the center of the condensate, we expect that
the variational {\it ansatz} in Eq.~(\ref{eq:ansatzzerot}) offers
a quantitatively good description of the wave function of the
condensate in the weakly-interacting limit that is determined by
the condition
\begin{equation}
   \frac{N_{\rm c}a}{\sqrt{\frac{\hbar}{m (\omega^2\omega_z)^{1/3}}}} \ll 1~,
\end{equation}
with $N_{\rm c}$ the number of atoms in the condensate.

To determine the equations for the variational parameters we note
that the Gross-Pitaevskii equation follows from a variation of the
action
\begin{eqnarray}
\label{eq:actionzerot}
 S[\phi^*,\phi] &=& \int\!dt\int\!d\bx \left\{
 \frac{i\hbar}{2} \left[ \phi^* \args \frac{\partial \phi \args}{\partial t}
 - \phi \args \frac{\partial \phi^* \args}{\partial t}\right]
 \right.
 \nonumber \\
 &+& \left. \phi^* \args \left[
 \frac{\hbar^2 {\bf \nabla}^2}{2m} - V^{\rm ext} (\bx) -
 \frac{T^{\rm 2B}}{2} |\phi \args|^2 \right] \phi \args \right\}~.
\end{eqnarray}
After substitution of the {\it ansatz} in
Eq.~(\ref{eq:ansatzzerot}) into this action we find the result
\begin{eqnarray}
\label{eq:zerotresult}
  S[{\bf u},q] &=& \int\!dt \left\{  \frac{N_{\rm c} \hbar}{q^2}
  \left[u_y (t) \dot u_x (t) - u_x (t) \dot u_y (t) +  \omega_{\rm p} (q,N_{\rm c}) u_y^2 (t)
        +\omega_{\rm p} (q,N_{\rm c}) u_x^2 (t)\right]
        \right. \nonumber \\
        && \ \ \ \ \ \ \left.
        -N_{\rm c} V(q,q_z,N_{\rm c}) + {\mathcal O} \left( u_i^3 \right) \rule{0mm}{5mm} \right\}~,
\end{eqnarray}
with the precession frequency of the vortex equal to
\begin{equation}
\label{eq:precfreq}
  \omega_{\rm p} (q,q_z,N_{\rm c}) =
  \frac{\hbar}{2mq^2}+\frac{m\omega^2q^2}{2\hbar}  - \frac{aN_{\rm
  c}\hbar}{\sqrt{2\pi}mq^2q_z}~,
\end{equation}
and the potential, which represents the total energy of the
condensate per particle, given by
\begin{equation}
\label{eq:pot}
  V(q,q_z,N_{\rm c}) =  \frac{\hbar^2}{m q^2} + \frac{\hbar^2}{4mq_z^2}
  +m \omega^2 q^2 +\frac{1}{4} m \omega_z q_z^2+
 \frac{aN_{\rm c}\hbar^2}{2\sqrt{2\pi}mq^2q_z}~.
\end{equation}
As a result we observe that the equilibrium widths of the
condensate in radial and axial direction, denoted from now on
again by by $q$ and $q_z$, respectively, are determined by
minimizing the potential $V(q,q_z,N_{\rm c})$ for a given number
of condensate atoms. By varying the action with respect to the
position of the vortex we find that the motion of the vortex is
determined by the equations
\begin{eqnarray}
\label{eq:eomzerot}
  \dot u_x (t) &=& -\omega_{\rm p} (N_{\rm c}) u_y
  (t)~, \nonumber \\
  \dot u_y (t) &=& \omega_{\rm p} (N_{\rm c}) u_x
  (t)~,
\end{eqnarray}
which imply that the vortex precesses around the center of the
condensate with precession frequency $\omega_{\rm p} (N_{\rm c})$,
which is equal to $\omega_{\rm p} (q,q_z,N_{\rm c})$ evaluated at
the equilibrium values of $q$ and $q_z$.

Although the above variational analysis has already provided us
with a simple description of the precession of the vortex around
the center of the condensate, it fails to account for the
experimental obervation that the vortex spirals out of the
condensate in the absence of rotation \cite{anderson2000}.
Contrary, the vortex remains, according the equations of motion in
Eq.~(\ref{eq:eomzerot}), at a fixed distance from the condensate
center. In particular, this implies that if the initial position
of the vortex is the center of the condensate, it will remain
there forever. This discrepancy between theory and experiment is
resolved by including the effects of thermal fluctuations, as we
will see in the next sections.

\section{Vortex dynamics at nonzero temperature}
\label{sec:finitet} The dynamics of a Bose-Einstein condensate is
at nonzero temperatures, i.e., in the presence of a thermal cloud
at temperature $T=1/(k_{\rm B} \beta)$ and with a chemical
potential $\mu$, determined by the stochastic Gross-Pitaevskii
equation \cite{stoof1997,stoof1999,stoof2001,duine2001}
\begin{eqnarray}
\label{eq:snlse}
   i \hbar \frac{\partial \phi (\bx, t)}{\partial t}&=&
   \left( 1+ \frac{\beta}{4} \hbar \Sigma^{\rm K} \args \right) \nonumber \\
     && \times \left\{ - \frac{\hbar^2 \nabla^2}{2m}
             + V^{\rm{ext}}(\bx) - \mu
         + T^{\rm 2B} |\phi \args|^2
     \right\}  \phi \args + \eta \args~.
\end{eqnarray}
This Langevin field equation quite generally generalizes the
Gross-Pitaevskii equation to nonzero temperatures, and includes
both dissipation, i.e., decay or growth of the Bose-Einstein
condensate, and (thermal) fluctuations. The dissipation and the
noise arise physically due to collisions between noncondensed
atoms in which one of the atoms is scattered into the condensate,
and the time-reversed process. The complex gaussian noise in the
stochastic Gross-Pitaevskii equation is completely determined by
the correlations
\begin{equation}
\label{eq:complexnoisecor}
  \langle \eta^* \args \eta (\bx',t') \rangle =
    \frac{i \hbar^2}{2}  \Sigma^{\rm K} \args
      \delta (t-t') \delta (\bx-\bx')~,
\end{equation}
where the strength of the noise is determined by the so-called
Keldysh self-energy, given by
\begin{eqnarray}
\label{eq:sigmak}
  \hbar \Sigma^{\rm K} \args & = & -4 \pi i \left( T^{\rm 2B}\right)^2
                           \int \frac{d \bk_1}{(2 \pi)^3}
                           \int \frac{d \bk_2}{(2 \pi)^3}
               \int \frac{d \bk_3}{(2 \pi)^3}
               (2 \pi)^3 \delta( \bk_1-\bk_2-\bk_3)
                                  \nonumber\\
           &   & \times \delta \left( \epsilon_1 -
                               \epsilon_2 -\epsilon_3 \right)
                 \left[ N_1 (1+N_2) (1+N_3) + (1+N_1) N_2 N_3 \right]~.
\end{eqnarray}
In this expression, $N_i$ is the Bose-distribution function of the
thermal cloud, evaluated at an energy of a thermal particle, which
is in the Hartree-Fock approximation given by
\begin{equation}
\label{energythermparttimedep}
  \epsilon_i = \frac{\hbar^2 \bk^2_i}{2 m} + V^{\rm{ext}} (\bx)
              +  2 T^{\rm 2B} |\langle \phi (\bx,t) \rangle|^2~.
\end{equation}
This approximation is valid in the regime where the temperature is
large compared to the zero-point energy of the external trapping
potential but smaller than the critical temperature. Note also
that we have taken the thermal cloud to be in equilibrium,
although this can easily be generalized.

In first approximation we neglect the inhomogeneity of the thermal
cloud over the size of the condensate. In good approximation, the
Keldysh self-energy then turns out to be given by
\cite{penckwitt2002}
\begin{equation}
\label{eq:resultsigmak}
  \hbar \Sigma^{\rm K} = \frac{-48 i m a^2 \left( k_{\rm B} T \right)^2}{\pi
  \hbar^2}~,
\end{equation}
which we will use from now on. Note that since the Keldysh
self-energy determines both the dissipation and the strength of
the fluctuations, the stochastic Gross-Pitaevskii equation can be
show to automatically fulfill the fluctuation-dissipation theorem
which ensures that the condensate relaxes to the correct
equilibrium distribution.

The (Wigner) probability distribution for the condensate wave
function, that results from the stochastic field equation in
Eq.~(\ref{eq:snlse}), can be written as the functional integral
\begin{equation}
\label{eq:probdistrcond}
  P[\phi, \phi^*; t] =
  \int^{\phi^* \args = \phi^* (\bx)}_{\phi \args = \phi (\bx)}
    d[\phi^*] d[\phi] \nonumber \\
    \exp \left\{ \frac{i}{\hbar} S^{\rm{eff}} [\phi^*,\phi]
    \right\}~,
\end{equation}
with the effective action given by
\begin{eqnarray}
\label{eq:actionfinitet}
  S^{\rm{eff}} [\phi^*, \phi] &=& \int_{t_0}^{t} dt' \int d \bx
  \frac{2}{\hbar \Sigma^{\rm K}}
     \left|
        \left(
      i \hbar \frac{\partial}{\partial t'} +
      \left\{ 1+\frac{\beta}{4}
      \hbar\Sigma^{\rm K}  \right\} \right. \right.  \nonumber \\
     && \ \ \ \ \times \left. \left. \left[ \frac{\hbar^2 \nabla^2}{2 m}
      -V^{\rm ext} (\bx) + \mu
      - T^{\rm 2B} |\phi (\bx,t')|^2 \right]
    \right) \phi (\bx,t')
     \right|^2~.
\end{eqnarray}
To determine the dynamics of the vortex at nonzero temperatures,
we have to slightly generalize the {\it ansatz} in
Eq.~(\ref{eq:ansatzzerot}) to allow also for fluctuations in the
number of condensate atoms and the global phase of the condensate
wave function \cite{duine2001}. Hence we now use the ansatz
\begin{eqnarray}
\label{eq:ansatzfinitet}
  \phi \args &=& \frac{\sqrt{N_{\rm c} (t) } e^{i \theta_0 (t)}}
  {\sqrt{\pi^{3/2} q^2 q_z \left[q^2+u_x^2 (t)+u^2_y(t)\right]}}
  \left\{\left[x-u_x(t)\right] +i \left[y-u_y(t)\right] \right\}
  \nonumber \\ && \times
  \exp \left\{ -\frac{x^2+y^2}{2 q^2} - \frac{z^2}{2
  q_z^2}\right\}~.
\end{eqnarray}
For the moment we consider the case $T^{\rm 2B}=0$, but keep the
Keldysh self-energy nonzero. This implies that we are dealing with
a noninteracting Bose gas in contact with a heat bath. Subsitution
of the above {\it ansatz} into the action leads to
\begin{eqnarray}
\label{eq:resultfinitet}
  S^{\rm eff} [N_{\rm c},\theta_{0},{\bf u}] &=& \int_{t_0}^t\!dt'
  \frac{2}{\hbar\Sigma^{\rm K}}
  \left\{ N_{\rm c} (t') \left[
  \hbar \frac{d\theta_0(t')}{dt'} + \mu_{\rm c} (t') +\frac{\hbar}{q^2}
  \left((u_x (t') \dot u_y (t') - u_y (t') \dot u_x (t')\right)
  -\mu \right]^2 \right. \nonumber \\
  &+& \frac{\hbar^2}{4 N_{\rm c} (t')}
  \left[ \frac{dN_{\rm c} (t')}{dt'}+\frac{\beta}{2}i\Sigma^{\rm K}
  \left( \mu_{\rm c} (t') -\mu \right) N_{\rm c} (t') \right]^2 \nonumber \\
  &+& \frac{\hbar^2 N_{\rm c}(t')}{q^2}
  \left[\dot u_x (t') + \omega_{\rm p} (q,0) u_y (t') - \gamma u_x
  (t')
  \right]^2 \nonumber \\
  &+& \left.  \frac{\hbar^2 N_{\rm c}(t')}{q^2}
  \left[ \dot u_y (t') - \omega_{\rm p} (q,0) u_x (t') - \gamma u_y
  (t')
  \right]^2 + {\mathcal O} \left( u_i^3\right)
  \rule{0mm}{6mm} \right\}~,
\end{eqnarray}
where we have again taken for the time-independent variational
parameters $q$ and $q_z$ the values obtained by minimizing the
potential in Eq.~(\ref{eq:pot}). The damping rate $\gamma$ is
given by
\begin{equation}
  \gamma = \frac{\beta \hbar^2 i \Sigma^{\rm K}}{8mq^2}~,
\end{equation}
and the condensate chemical potential reads
\begin{equation}
  \mu_{\rm c} (t) = \frac{\partial (N_{\rm c} V(q,q_z,0))}{\partial N_{\rm c}}
     + \frac{\hbar^2}{2mq^4} \left[ u_x^2 (t) + u_y^2 (t) \right]~.
\end{equation}

From the action for the variational parameters we are able to
deduce the stochastic rate equation for the number of atoms and
the stochastic equations of motion for the global phase and for
the position of the vortex. First, the stochastic rate equation
for the number of atoms is given by \cite{duine2001}
\begin{equation}
  \frac{dN_{\rm c} (t)}{dt} = -\frac{\beta}{2} i \Sigma^{\rm K} \left[
  \mu_{\rm c} (t) - \mu\right] N_{\rm c} (t) + 2 \sqrt{N_{\rm c} (t)}
  \eta (t)~,
\end{equation}
with the correlations of the Gaussian noise given by
\begin{equation}
  \langle \eta (t') \eta (t) \rangle = \frac{i\Sigma^{\rm K}}{4}
   \delta (t-t')~.
\end{equation}
Second, the stochastic equation of motion for $\theta_0(t)$ is
given by \cite{duine2001}
\begin{equation}
  \hbar \frac{d\theta_0(t)}{dt} = \mu - \mu_{\rm c} (t) +
  \frac{\nu (t)}{\sqrt{N_{\rm c} (t)}}~,
\end{equation}
where the correlation of the gaussian noise are given by
\begin{equation}
  \langle \nu (t') \nu (t) \rangle = \frac{i \hbar^2\Sigma^{\rm K}}{4}
   \delta (t-t')~.
\end{equation}

Since we are mostly interested in the stochastic equations of
motion for the position of the vortex, we neglect from now on the
fluctuations in the number of condensate atoms and the global
phase of the condensate. The Langevin equations for the position
of the vortex core are, from the action in
Eq.~(\ref{eq:resultfinitet}), seen to be given by
\begin{eqnarray}
\label{eq:eomfinitet}
  \dot u_x (t) &=&-\omega_{\rm p} (0) u_y (t)+ \gamma u_x
  (t) + \eta_x (t)~, \nonumber \\
  \dot u_y (t) &=& \omega_{\rm p} (0) u_x
  (t) +\gamma u_y (t) + \eta_y (t)~,
\end{eqnarray}
where the gaussian noise is completely determined by
\begin{eqnarray}
\label{eq:corfuncnoise}
  \langle \eta_i (t) \eta_j (t')\rangle = \frac{q^2 i \Sigma^{\rm K}}{4 N_{\rm
  c}} \delta_{ij} \delta (t-t') \equiv \sigma \delta_{ij} \delta (t-t')~.
\end{eqnarray}
The stochastic equations for the position of the vortex core in
Eq.~(\ref{eq:eomfinitet}) have two extra terms with respect to the
zero-temperature result in Eq.~(\ref{eq:eomzerot}). First, the
stochastic equations have a damping term proportional to the
damping rate $\gamma$. As we will see, this term leads to
exponentional increase in the distance between the vortex position
and the center of the condensate. Secondly, the stochastic
equations have an additive noise term which represents the effect
of thermal fluctuations on the position of the vortex core. If
these noise terms would not have been present, the lifetime of the
vortex would again be infinite if the initial position of the
vortex is the center of the condensate. Physically, the vortex is
``kicked'' out of the center of the condensate due to thermal
fluctuations which are represented by the gaussian noise terms in
Eq.~(\ref{eq:eomfinitet}).

Although we have in first instance derived the stochastic equation
for the position of the vortex core by assuming that we are
dealing with a Bose-Einstein condensate of noninteracting atoms,
we can now consider the interacting case by replacing the
precession frequency in $\omega_{\rm p} (0)$ by $\omega_{\rm p}
(N_{\rm c})$. The Fokker-Planck equation, that determines the
probability distribution for the position of the vortex, is then
given by
\begin{eqnarray}
\label{eq:fpeqnu}
  \frac{\partial P[u_x,u_y,t]}{\partial t}
  &=& \left\{ \frac{\partial}{\partial u_x}
       \left[ \omega_{\rm p} (N_{\rm c}) u_y - \gamma u_x  \right]
       + \frac{\partial}{\partial u_y}
       \left[ -\omega_{\rm p} (N_{\rm c}) u_x - \gamma u_y  \right]
       \right. \nonumber\\ && \left.
       + \frac{\sigma}{2} \left[ \frac{\partial^2}{\partial u_x^2}
       + \frac{\partial^2}{\partial u_y^2} \right]
  \right\} P[u_x,u_y,t]~.
\end{eqnarray}
In the next section, we use this equation to determine the average
distance of the vortex to the center of the condensate. This
result is then used to obtain the lifetime of the vortex.

\section{Lifetime of the vortex}
\label{sec:lifetime} The average distance of the vortex to the
center of the condensate is given by
\begin{equation}
  r (t) \equiv \left\langle \sqrt{u_x^2+u_y^2} \right\rangle (t) \equiv
   \int du_x du_y \sqrt{u_x^2+u_y^2}~P[u_x,u_y,t]~.
\end{equation}
With the Fokker-Planck equation for the position of the vortex
core in Eq.~(\ref{eq:fpeqnu}) we find, with the use of a partial
integration, that $r(t)$ obeys the equation of motion
\begin{equation}
\label{eq:eomraverage}
  \frac{dr(t)}{dt} = \gamma r(t) + \frac{\sigma}{2r(t)}~.
\end{equation}
The general solution of this equation is given by
\begin{equation}
\label{eq:solutionaverage}
  r(t) = \frac{\sqrt{\left[2\gamma r^2(0)+\sigma\right]e^{2 \gamma t} - \sigma}}{\sqrt{2
  \gamma}}~.
\end{equation}
Note that for small times $t$ and distance $r(t)$ we have that $r
(t) \propto \sqrt{t}$. This may be understood from the fact that,
if we neglect the contribution of the precession frequency
$\omega_{\rm p} (N_{\rm c})$ and the damping rate $\gamma$ to the
Fokker-Planck equation in Eq.~(\ref{eq:fpeqnu}), the vortex simply
undergoes Brownian motion. For large times $t$ and distance $r(t)$
we have that $r(t) \propto e^{\gamma t}$, as is easily seen from
Eqs.~(\ref{eq:eomraverage})~and~(\ref{eq:solutionaverage}).

For a given initial position $r(0)= r_{\rm min}$, the vortex
lifetime is defined as the time it takes the vortex to reach the
edge of the condensate at $r_{\rm max}$. Of course, for our
gaussian {\it ansatz} the edge of the condensate is not defined
uniquely, but we could, for instance, take $r_{\rm max}=q$. In the
Thomas-Fermi limit we have $r_{\rm max} =R$, with $R$ the
Thomas-Fermi radius. The lifetime of the vortex is in first
instance given by
\begin{equation}
\label{eq:lifetimegeneral}
  \tau = \frac{1}{2\gamma} \ln
  \left [ \frac{r_{\rm max}^2+ \sigma/(2\gamma)}{r_{\rm
  min}^2+\sigma/(2\gamma)}\right]~.
\end{equation}
Let us first discuss the case where we neglect the contribution of
the noise due to thermal fluctuations, i.e., we take $\sigma=0$.
The lifetime of the vortex is then given by
\begin{equation}
\label{eq:taunonoise}
  \tau_0 = \frac{\pi}{6} \left( \frac{q}{a}\right)^2 \frac{\hbar}{\left( k_{\rm B} T \right)}
    \ln \left( \frac{r_{\rm max}}{r_{\rm min}}\right)~.
\end{equation}
This result is very similar to the result obtained by Fedichev and
Shlyapnikov \cite{fedichev1999} who indeed neglect thermal
fluctuations. Schmidt {\it et al.} \cite{schmidt2003}, who include
fluctuations, numerically find a similar result. In particular,
the dependence on the ratio $r_{\rm max }/r_{\rm min}$ is
identical, and we observe that, in the absence of thermal
fluctuations, the lifetime of the vortex diverges as $r_{\rm min}
\to 0$. Interestingly, although we consider the weakly-interacting
limit, as opposed to Fedichev and Schlyapnikov who consider the
Thomas-Fermi limit, the temperature dependence of the prefactor of
the lifetime in Eq.~(\ref{eq:taunonoise}) is the same in both
results. This is even more surprising because of the fact that the
dissipation of the vortex considered by Fedichev and Shlyapnikov
is physically due to scattering of quasiparticles of the vortex
core. This mechanism of decay is physically quite distinct from to
the collisional effects we consider here.

Including thermal fluctuations, the vortex lifetime is finite even
if its initial position is the center of the condensate. In
particular, for that case we have that
\begin{equation}
  \tau = \frac{\pi}{12} \left( \frac{q}{a}\right)^2 \frac{\hbar}{\left( k_{\rm B} T \right)}
    \ln \left[ 1 + \frac{N_{\rm c} \hbar^2 r_{\rm max}^2}{mq^4} \frac{1}{\left( k_{\rm B} T
    \right)}\right]~.
\end{equation}
We observe that the leading low-temperature dependence of the
lifetime of the vortex is $T^{-1} \ln \left[ N_{\rm c} \hbar
\omega/(k_{\rm B}T) \right]$, whereas in the case where we
neglected the thermal fluctuations we had that the lifetime was
proportional to $T^{-1}$.

\section{Conclusions}
\label{sec:conclusions} We have studied, by means of a variational
analysis, the motion of a vortex in a Bose-Einstein condensate.
The main results were the stochastic equations of motion for the
position of the vortex and their corresponding Fokker-Planck
equation, which enabled us to analytically calculate the lifetime
of the vortex. We have compared our results for the latter with
the available theoretical results \cite{fedichev1999,schmidt2003}.
It is surprising that, although the collisional decay mechanism
that is considered here is physically quite different from the
decay mechanism discussed by Fedichev and Shlyapnikov
\cite{fedichev1999}, we nevertheless find the same temperature
dependence if we neglect the thermal fluctuations. Including
thermal fluctuations leads to a different temperature dependence.
Note also that, because we consider condensate dissipation due to
collisions, the lifetime of the vortex is inversely proportional
to the square of the scattering length. The result for the
lifetime found by Fedichev and Schlyapnikov is proportional to
$a^{-4/5}$. It would therefore be very interesting to
experimenally measure the dependence on the temperature and the
scattering length of the vortex lifetime.

Our present analysis was limited to the case where the overall
density profile is close to a gaussian, i.e., the so-called
weakly-interacting limit. In principle, our treatment can be
generalized to the case where the condensate is described by a
Thomas-Fermi density profile. However, in this limit technical
problems occur due to the fact that we need to incorporate the
density profile of the core of the vortex. This strongly limits
the feasibility of analytical results. We expect, however, that
the main difference between the weakly-interacting limit and the
Thomas-Fermi limit is the difference in the precession frequency
of the vortex, which can easily be found from a zero-temperature
approach \cite{fetter2001}. Therefore, we believe that the
lifetime results presented here provide also in the Thomas-Fermi
limit a more than qualitative understanding of the dynamics of a
vortex at nonzero temperatures.

\acknowledgments It is a pleasure to thank Jani Martikainen for
helpful remarks.

\end{document}